\documentclass[preprint,showpacs,preprintnumbers,amsmath,amssymb]{revtex4}

\usepackage{graphicx}
\usepackage{dcolumn}
\usepackage{bm}
\usepackage{float,color}

\newcommand{\bff}[1]{{\mbox{\boldmath $#1$}}}

\begin{document}

\title{
An efficient method for computing the Thouless-Valatin inertia parameters}

\author{Z. P. Li$^{1,2}$, T. Nik\v{s}i\'{c}$^2$, P. Ring$^3$, D. Vretenar$^2$, J. M. Yao$^1$, and J. Meng$^{4,5}$}
\affiliation{$^{1}$School of Physical Science and Technology, Southwest University, Chongqing 400715, China}
\affiliation{$^{2}$Physics Department, Faculty of Science, University of Zagreb, 10000 Zagreb, Croatia}
\affiliation{$^{3}$Physik-Department der Technischen Universit\"at M\"unchen, D-85748 Garching, Germany}
\affiliation{$^{4}$State Key Laboratory of Nuclear Physics and Technology, School of Physics,
Peking University, Beijing 100871, China}
\affiliation{$^{5}$School of Physics and Nuclear Energy Engineering, Beihang University, Beijing 100191, China}


\begin{abstract}
Starting from the adiabatic time-dependent Hartree-Fock approximation (ATDHF),
we propose an  efficient method to calculate the Thouless-Valatin moments of inertia for
the nuclear system. The method is based on the rapid convergence of the expansion of the inertia matrix.
The accuracy of the proposed method is verified in the rotational case by comparing the results with the
exact Thouless-Valatin moments of inertia calculated using the self-consistent cranking model.
The proposed method is computationally much more efficient than the  full ATDHF calculation,
yet it retains a high accuracy of the order of $1$\%. 
\end{abstract}

\date{\today}
\maketitle

\section{\label{sec-introduction}Introduction}
The variation of nuclear ground-state shapes is governed by the modification of the shell-structure
of single-nucleon orbitals. Far from the valley of $\beta$-stability, the energy spacings
between single-nucleon levels change considerably with the number of neutrons and/or protons.
The reduction of spherical shell closure is often associated with the occurrence of deformed
ground states and, in many cases, with the phenomenon of coexistence of different shapes in a
single nucleus.  A quantitative description of the evolution of nuclear shapes, including regions
of short-lived exotic nuclei that are becoming accessible in experiments at radioactive-beam
facilities, necessitate accurate modeling of the underlying microscopic nucleonic dynamics.
Major advances in nuclear theory have recently been made in studies of complex shapes
and the corresponding excitation spectra and electromagnetic decay patterns, especially in
the framework of nuclear energy density functionals (EDFs)~\cite{EDF.04,BHR.03,VALR.05,Meng.06,Dob.11}.

A microscopic, EDF-based description of complex collective excitation spectra usually starts
from a constrained Hartree-Fock plus BCS (HFBCS) or Hartree-Fock-Bogoliubov (HFB)
calculation of the binding energy surface with the mass multipole moments as constrained
quantities. The static nuclear mean-field is characterized by symmetry breaking: translational,
rotational and particle number. Even though symmetry breaking incorporates important static
correlations (e.g., deformations and pairing), the static self-consistent solution can only provide
an approximate description of bulk ground-state properties such as masses and radii. Modeling
excitation spectra and transition rates in the EDF framework necessitates a systematic treatment
of dynamical effects related to restoration of broken symmetries and fluctuations in collective coordinates.

One possible approach to five-dimensional quadrupole dynamics that restores rotational symmetry
and allows for fluctuations around triaxial mean-field minima is to formulate a collective Hamiltonian,
with deformation-dependent inertia parameters determined by microscopic self-consistent
mean-field calculations.
The dynamics of the collective Bohr Hamiltonian is governed by the vibrational inertial
functions and the moments of inertia~\cite{GR.80}. For these quantities either the Gaussian overlap approximation of 
the generator coordinate method (GCM-GOA) (Yoccoz masses~\cite{Yocc.57}) or the adiabatic time-dependent Hartree-Fock-Bogoliubov (ATDHFB)
expressions (Thouless-Valatin masses~\cite{TV.62}) can be used. The Thouless-Valatin
masses have the advantage that they also include the time-odd
components of the self-consistent mean field and,
in this sense, the full dynamics of a nuclear system. This can be seen most clearly in the case of
translational motion, where the Thouless-Valatin mass corresponds to the
exact mass $A\cdot m$ of $A$ nucleons~\cite{Vi.63}, whereas the GCM method
produces the exact value only when the center of the mass
velocity is also included as the generator coordinate~\cite{RY.66}. The calculation of the
Thouless-Valatin masses is often simplified by adopting the cranking formulas~\cite{GG.79,Ing.56}
that neglect the residual interaction. In that case the Thouless-Valatin corrections are usually
taken into account by scaling the inertia parameters with an empirical
factor ($\approx 1.2 - 1.4$)~\cite{DS.81,LGD.99,Pro.04}.

In this work we present an efficient method to calculate the Thouless-Valatin moments of inertia for
the nuclear system.  The method is based on the rapid convergence of the expansion of the inertia matrix.
The accuracy of the proposed method is verified in the rotational case by comparing the results with the
exact Thouless-Valatin moments of inertia calculated using the self-consistent cranking model.
The proposed method is computationally much more efficient than the  full ATDHF calculation,
yet it retains a high accuracy of the order of $1$\%.

\section{\label{sec-theoretical}Theoretical framework}
We begin with a brief review of the adiabatic time-dependent Hartree-Fock
theory. A more detailed exposition of this formalism can be found, for instance,
in Refs.~\cite{BV.78,RS.80}.
The aim of the ATDHF theory is to derive in a fully microscopic and
consistent way a Hamiltonian for the description of collective phenomena in which many nucleons act
coherently. The theory is based on two approximations: i) in the TDHF one assumes that
the many-body time-dependent wave function of the system is
a Slater determinant at all times; and ii) in the
adiabatic approximation the collective motion is slow compared to single-particle motion
and, therefore, the collective kinetic energy is a quadratic function of the velocities.

To identify the components of the density matrix that correspond to the coordinates and momenta of
the collective Hamiltonian, we recall that the coordinates are even and the momenta are odd
under time-reversal, and decompose the density matrix in the following way:
\begin{equation}
\label{eq:decomposition}
\rho(t) = e^{i\chi(t)/\hbar}\rho_0(t) e^{-i\chi(t)/\hbar}.
\end{equation}
Both matrices, $\rho_0(t)$ and $\chi(t)$, are Hermitian and time-even. $\rho_0(t)$ represents the
coordinates of the collective Hamiltonian,
and $\chi(t)$ is the ``adiabaticity parameter" that must be small compared to unity.
At all times $\rho(t)$ is a Slater determinant, that is, $\rho_0(t)^2=\rho_0(t)$
and ${\rm Tr} \rho_0=N$, $N$ being the particle number.
In the following we work in the basis in which $\rho_0$ is diagonal, and consequently
the operators $\rho_0$ and $\sigma_0=1-\rho_0$  project onto hole and particle
states, respectively. This basis depends on time because $\rho_0(t)$ is a function of time.

In the adiabatic approximation it is assumed that the total density $\rho(t)$ of the system
is always close to the density $\rho_0(t)$, that is, the matrix $\chi$ that introduces the time-odd
components remains small at all times. Expanding the density matrix to
second order in the operator $\chi$, the following expression is obtained:
\begin{equation}
\label{eq:rho-expansion}
\rho \approx \left( 1+\frac{i}{\hbar}\chi -\frac{1}{\hbar^2}\chi^2 \right)\rho_0
\left( 1-\frac{i}{\hbar}\chi -\frac{1}{\hbar^2}\chi^2 \right)\approx \rho_0 +\rho_1+\rho_2,
\end{equation}
where
\begin{align}
\label{eq:rho1}
\rho_1&=\frac{i}{\hbar}\left[\chi,\rho_0\right]=\frac{i}{\hbar}\left(\chi\rho_0-\rho_0\chi \right),\\
\label{eq:rho2}
\rho_2&=-\frac{1}{2\hbar^2}\left[\chi,\left[\chi,\rho_0\right]\right] =
         \frac{1}{\hbar^2} \chi \rho_0 \chi - \frac{1}{2\hbar^2}\left( \chi^2\rho_0 +\rho_0\chi^2\right) .
\end{align}
$\rho_1$ is linear in $\chi$, time-odd, and has only $ph$ and $hp$ non-vanishing matrix elements.
$\rho_2$ is quadratic in $\chi$, therefore time-even, and has only $hh$ and
$pp$ matrix elements. The many-body Hamiltonian can also be expanded to second order in the
operator $\chi$:
\begin{align}
h_{ab}(\rho) &= t_{ab} + \sum_{cd}{V_{adbc}\rho_{cd}}
= t_{ab} + \sum_{cd}{V_{adbc}(\rho_0)_{cd}}+\sum_{cd}{V_{adbc}(\rho_1)_{cd}}
+\sum_{cd}{V_{adbc}(\rho_2)_{cd}} ,
\end{align}
where $t$ is the kinetic energy operator, and $V$ denotes a generic two-body interaction.
The Hamiltonian contains time-even ($h_0$ and $\Gamma_2$) and time-odd parts ($\Gamma_1$)
\begin{equation}
h(\rho) = h_0 + \Gamma_1 + \Gamma_2.
\end{equation}
Consequently the time-dependent Hartree-Fock equation $i\hbar \dot{\rho}=[h,\rho]$ also decomposes
into two equations:
\begin{align}
\label{eq:dot-rho-0}
i\hbar \dot{\rho}_0 &= [h_0,\rho_1] + [\Gamma_1,\rho_0], \\
\label{eq:dot-rho-1}
i\hbar \dot{\rho}_1 &= [h_0,\rho_0]+[\Gamma_1,\rho_1] + [\Gamma_2,\rho_0].
\end{align}
In Eq.~(\ref{eq:dot-rho-1}) the term $[h_0, \rho_2]$ has been neglected because the
$ph$ and $hp$ components are small, and the $pp$ and $hh$ parts vanish \cite{RS.80}.
The total energy of the system
\begin{equation}
E=\sum_{ab}{t_{ab}\rho_{ba}} + \frac{1}{2}\sum_{abcd}{ \rho_{ba} V_{adbc}\rho_{cd} },
\end{equation}
can be expressed in terms of the variables $\rho_0$ and $\rho_1$, or $\rho_0$ and $\chi$.
Terms which depend on the velocity in second order build the kinetic energy of the collective
Hamiltonian:
\begin{equation}
K = \sum_{ab}{(h_0)_{ab}(\rho_2)_{ba}} +  \frac{1}{2}\sum_{abcd}{ (\rho_1)_{ba} V_{adbc}(\rho_1)_{cd} }.
\label{E10}
\end{equation}
We recall that the matrix $\rho_0$ projects onto the hole states and, inserting
Eqs. (\ref{eq:rho1}) and (\ref{eq:rho2}) in the expression above, the kinetic energy can be written:
\begin{equation}
K = \frac{1}{2\hbar^2} \left( \begin{array}{cc} \chi^* & \chi\end{array}\right)
\left( \begin{array}{cc} A & -B \\ -B^* & A \end{array} \right)
\left( \begin{array}{c} \chi \\ \chi^* \end{array} \right)\; ,
\end{equation}
where the matrix $A$ is Hermitian, and $B$ is symmetric
\begin{equation}
\label{eq:AB}
A_{php^\prime h^\prime} = (\epsilon_p - \epsilon_h)\delta_{pp^\prime} \delta_{hh^\prime}
+V_{phh^\prime p^\prime}, \quad
B_{php^\prime h^\prime} = V_{php^\prime h^\prime}.
\end{equation}

In Ref.~\cite{DR.09} it is shown that the effective $ph$-interaction in relativistic point coupling models can be written as
a sum of separable terms
\begin{equation}
V_{adbc}^{}=\sum_{r}Q_{ab}^{r}V_{r}Q_{cd}^{r\ast}
\label{separable}%
\end{equation}
The same is true for relativistic Hartree models with meson exchange forces. The single particle operators $Q^{r}$ are either even or
odd under time-reversal. The time-odd operators correspond to the isoscalar and isovector currents ${\bm j}({\bm r})$ and
$\vec{\tau}{\bm j}({\bm r})$. Implementing the interaction (\ref{separable}) into Eq. (\ref{E10}) we find
\begin{equation}
\frac{1}{2}\sum_{abcd}{ (\rho_1)_{ba} V_{adbc}(\rho_1)_{cd} }=
\sum_{r}%
\mathrm{Tr}(Q^{r}\rho^{}_{1})V_{r}\mathrm{Tr}(Q^{r}\rho_{1}^{})^{*}.%
\end{equation}
Since $\rho_{1}$ is time-odd the traces vanish for time-even operators and only the time-odd operators in the
matrices $A$ and $B$ contribute to the inertia parameters.

The equation of motion (\ref{eq:dot-rho-0}) can be written into the following form:
\begin{equation}
\left( \begin{array}{c} \dot{\rho}_0 \\ \dot{\rho}_0^* \end{array} \right)
=\frac{1}{\hbar^2}
\left( \begin{array}{cc} A  & -B \\ -B^* & A^*  \end{array} \right)
\left( \begin{array}{c} \chi \\ \chi^* \end{array} \right)
\equiv \mathcal{M}^{-1} \left( \begin{array}{c} \chi \\ \chi^* \end{array} \right) \;.
\label{E13}
\end{equation}

To perform realistic calculations the dimension of the problem has to be reduced,
that is, one has to select a small number of active degrees of freedom $q_1,\dots, q_f$.
This means that we are able to generate a subset of time-even Slater determinants,
characterized by the parameters $\mathbf{q}$, with the following property: the solution of the ATDHF
problem will always remain within this subset of Slater determinants.
In other words, we have found a path $\rho_0(t) = \rho_0[\mathbf{q}(t)]$
from which we can calculate the velocity
\begin{equation}
\dot{\rho}_0(t)= \dot{\mathbf{q}}(t) \frac{\partial \rho_0}{\partial \mathbf{q}}.
\end{equation}
Next we define the operator $\mathbf{P}$ with the relation:
\begin{equation}
\frac{\partial \rho_0}{\partial \mathbf{q}} = -\frac{i}{\hbar} [\mathbf{P},\rho_0] \; ,
\end{equation}
and obtain the following expression for the kinetic energy:
\begin{equation}
K=\frac{1}{2} \sum_{\mu,\nu=1}^f{M_{\mu\nu}(\mathbf{q})\dot{q}_\mu\dot{q}_{\nu}  }\; ,
\end{equation}
where $M_{\mu \mu^\prime}(\mathbf{q})$ denotes the real collective mass tensor
\begin{equation}
M_{\mu \mu^\prime}(\mathbf{q}) =
\frac{1}{\hbar^2} \left(\begin{array}{cc} P^* & -P\end{array}\right)_\mu
\mathcal{M} \left( \begin{array}{c} P \\ -P^*\end{array} \right)_{\mu^\prime}.
\label{eq:E17}
\end{equation}

To evaluate $\mathcal{M}$, we have to invert the matrix
\begin{equation}
\mathcal{M}^{-1} =\left( \begin{array}{cc} A  & -B \\ -B^* & A^*  \end{array} \right) = \mathcal{M}^{-1}_0 + \mathcal{V}
\label{eq:E18}
\end{equation}
in Eq. (\ref{E13}). For this purpose we decompose the matrix into a diagonal part containing the energies
of particle and hole states
\begin{equation}
\left(\mathcal{M}^{-1}_0\right)_{php'h'} = (\epsilon_p - \epsilon_h)\delta_{pp^\prime} \delta_{hh^\prime} \;,
\label{eq:M_0}
\end{equation}
and the residual interaction $\mathcal{V}$, and use the fact that the interaction matrix elements
are in most cases much smaller than the $ph$-energies. This is because only the time-odd components
of the residual interaction contribute. Therefore the matrix $\mathcal{M}$ can be written in the following form:
\begin{equation}
\mathcal{M} = \left[ \mathcal{M}^{-1}_0 +  \mathcal{V}\right]^{-1}
=\mathcal{M}_0\left[\openone +  \mathcal{V}\mathcal{M}_0\right]^{-1}
\end{equation}
We expand the factor in the square bracket and obtain:
%
%
\begin{equation}
\label{eq:M-expansion}
\mathcal{M} = \mathcal{M}_0 -  \mathcal{M}_0 \mathcal{V}\mathcal{M}_0
+ \mathcal{M}_0 \mathcal{V} \mathcal{M}_0 \mathcal{V}
\mathcal{M}_0+\cdots.
\end{equation}
Since $\mathcal{M}^{-1}_0$ is diagonal, inverting this matrix is trivial and the problem is
reduced to simple matrix multiplications.
The zero-order term, of course, yields the Inglis-Belyaev formula:
%
%
%
\begin{equation}
M^0_{\mu \mu^\prime}(\mathbf{q}) = \frac{1}{\hbar^2}
 \left(\begin{array}{cc} P^* & -P\end{array}\right)_\mu
\mathcal{M}_0 \left( \begin{array}{c} P \\ -P^*\end{array} \right)_{\mu^\prime}
=\frac{2}{\hbar^2}\sum_{ph}{\frac{|\hat{P}_{ph}|^2}{\epsilon_p-\epsilon_h} } .
\end{equation}
The first- and the second-order terms
%
%
\begin{equation}
\mathcal{M}_1 = -  \mathcal{M}_0 \mathcal{V}\mathcal{M}_0,\quad
\mathcal{M}_2 =  \mathcal{M}_0 \mathcal{V} \mathcal{M}_0\mathcal{V}\mathcal{M}_0.
\label{E23}
\end{equation}
represent the leading corrections to the Inglis-Belyaev formula. The purpose of this exploratory study
is to determine the convergence of the expansion (\ref{eq:M-expansion}), as well as the level
of agreement with the Thouless-Valatin formula. In this work
we only consider the moments of inertia for collective rotation,
that is, the operator $\hat{P}$ corresponds to the components of the angular momentum vector .

In fact, for a stationary deformed solution without external
constraint, as it is discussed in the following application, the RPA-equation
has a Goldstone mode. As discussed in detail in Sect. 8.4.7 of Ref.~\cite{RS.80},
from rotational invariance of the Hamiltonian, i.e. from $[H,J_{x}]=0$ we obtain
for $P=J_{x}$ a spurious solution%
\begin{equation}
\mathcal{M}^{-1}\left(
\begin{array}
[c]{c}%
P\\
P^{\ast}%
\end{array}
\right)  =\left(
\begin{array}
[c]{cc}%
A & -B\\
-B^{\ast} & A^{\ast}%
\end{array}
\right)  \left(
\begin{array}
[c]{c}%
P\\
P^{\ast}%
\end{array}
\right)  =0\label{goldstone}%
\end{equation}
On the other side, from Eq. (\ref{eq:E17}) we see that we have to solve the inhomogeneous equation%
\begin{equation}
\left(
\begin{array}
[c]{cc}%
A & -B\\
-B^{\ast} & A^{\ast}%
\end{array}
\right)  \left(
\begin{array}
[c]{c}%
X\\
Y
\end{array}
\right)  =\left(
\begin{array}
[c]{c}%
P\\
-P^{\ast}%
\end{array}
\right)  .
\end{equation}
Such a solution exists, because the inhomogeneous part of this equation
is orthogonal to the Goldstone mode of Eq. (\ref{goldstone}).
Of course, the explicit inversion of the matrix $\mathcal{M}^{-1}$
is technically complicated because it has to be carried out in the space
orthogonal to the Goldstone mode. However, the method proposed here avoids
these technical complications. In each order of the approximation the matrix
(\ref{eq:M-expansion}) acts on the vector
\begin{equation}
\left(
\begin{array}
[c]{c}%
P\\
-P^{\ast}%
\end{array}
\right)
\end{equation}
which eliminates all spurious contributions. We also have to
emphasize, that the problem of the Goldstone mode occurs only at the stationary
points of the energy surface, where the constraint vanishes. For all other solutions
the constraining operator does not commute with the angular momentum and therefore there
exist no spurious solution.


As a specific example of the nuclear energy density functional
we consider the point-coupling implementation of a relativistic EDF -- the functional PC-F1~\cite{Bur.02}:
\begin{align}
E_{\textrm{RMF}} &= \int{d\mathbf{r}\mathcal{E}_{\textrm{RMF}} (\mathbf{r}) }
  = \sum_{i=1}^A{\int{d\mathbf{r} \bar{\psi}_i(\mathbf{r}) (-i\mathbf{\gamma}\mathbf{\nabla}+m)
                      \psi_i(\mathbf{r})   }  } \nonumber \\
  &+ \int d\mathbf{r} \left[\frac{1}{2}\alpha_S\rho_S^2 + \frac{1}{3}\beta_S \rho_S^3
    +  \frac{1}{4}\beta_S \rho_S^4 + \delta_S \rho_S \triangle \rho_S
     +  \frac{1}{2}\alpha_Vj_\mu j^\mu +  \frac{1}{4}\gamma_V(j_\mu j^\mu)^2
     +  \frac{1}{2}\delta_V j_\mu \Delta j^\mu \right. \nonumber \\
     &\left. +  \frac{1}{2} \alpha_{TV} (j_{TV})_\mu (j_{TV})^\mu
     +  \frac{1}{2} \delta_{TV} (j_{TV})_\mu \Delta (j_{TV})^\mu
     +  \frac{1}{2} \alpha_{TS} \rho_{TS}^2
     +  \frac{1}{2} \delta_{TS}  \rho_{TS} \Delta  \rho_{TS} +  \frac{e}{2}\rho_p A^0
      \right]  ,
\label{eq:E_RMF}
\end{align}
where $\psi$ denotes the Dirac spinor field of a nucleon, and the local isoscalar and
isovector densities and currents
\begin{align}
\rho_S(\mathbf{r}) &= \sum_{i=1}^A{\bar{\psi}_i(\mathbf{r})\psi_i(\mathbf{r})  },\\
\rho_{TS}(\mathbf{r}) &= \sum_{i=1}^A{\bar{\psi}_i(\mathbf{r})\tau_3 \psi_i(\mathbf{r})  },\\
j^{\mu}(\mathbf{r}) &= \sum_{i=1}^A{\bar{\psi}_i(\mathbf{r})\gamma^\mu \psi_i(\mathbf{r})  }, \\
j_{TV}^{\mu}(\mathbf{r}) &= \sum_{i=1}^A{\bar{\psi}_i(\mathbf{r})\gamma^\mu\tau_3 \psi_i(\mathbf{r})  },
\end{align}
are calculated in the {\it no-sea} approximation: the summation runs over all occupied states in the
Fermi sea. This means that only occupied single-nucleon states with positive energy explicitly contribute to
the nucleon self-energies. In Eq. (\ref{eq:E_RMF}) $\rho_p$ is the proton density, and $A^0$
denotes the Coulomb potential.

The matrix elements of the residual interaction are derived from the EDF Eq. (\ref{eq:E_RMF})
\begin{equation}
V_{adbc} = \frac{\partial^2 E_{RMF}}{\partial \rho_{ba}\partial \rho_{cd}},
\end{equation}
where generic indices $(a,b,c,d,\dots)$ denote quantum numbers that specify the single-nucleon
state $\{\psi_a\}$. These belong to three distinct sets: the index $p$ (particle) denotes unoccupied
states above the Fermi sea, the index $h$ (hole) is for occupied states in the Fermi sea, and
with $\alpha$ we denote the unoccupied negative-energy states in the Dirac sea.
The calculation of the moments of inertia involves only the time-odd terms of the residual
interaction, for which the isoscalar-vector channel plays the dominant role.
The time-odd contributions of the isovector-vector and the  electromagnetic fields are omitted because the corresponding couplings are small in comparison to the isoscalar-vector coupling.
Here we make a further simplification by assuming
that the nonlinear and the derivative terms can be neglected, that is, it is sufficient to
retain only the linear isoscalar-vector term (see Fig.~\ref{fig:Sm154-cranking} and Tab.~\ref{Tab:Sm154-craning}):
\begin{equation}
V_{adbc} = -\alpha_V
\int{\ [ \psi_a^\dagger \bff{\alpha} \psi^{}_b] [  \psi_d^\dagger \bff{\alpha} \psi^{}_c ] d^3 r  }\; .
\end{equation}

\section{\label{sec-numerical-tests} Numerical test}

\begin{figure}[htb!]
\centering
\includegraphics[scale=0.4]{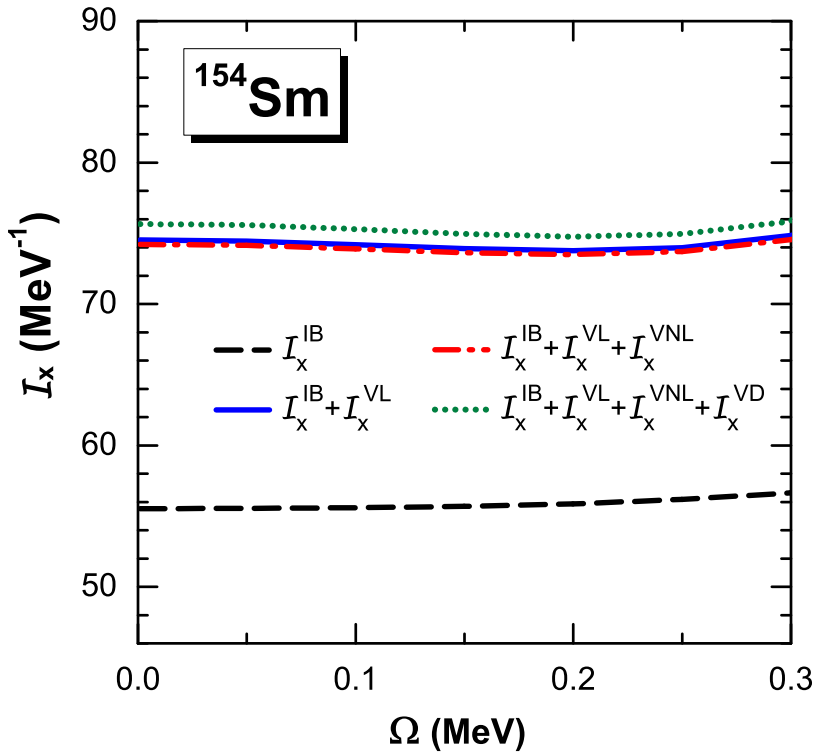}
\caption{\label{fig:Sm154-cranking}(Color online)
The moment of inertia ${\cal{I}}_x$ of $^{154}$Sm computed using the cranking RMF framework.
The various curves correspond to calculations that successively include the following terms
of the residual interaction:
the time-even part (Inglis-Belyaev formula denoted by IB), the linear (VL),
nonlinear (VNL), and derivative (VD) time-odd terms in the isoscalar-vector channel. }
\end{figure}
\begin{table}[htb!]
\tabcolsep=6pt
\caption{\label{Tab:Sm154-craning}
Contributions to the moment of inertia ${\cal{I}}_x$ of different terms of the residual interaction:
the time-even part (Inglis-Belyaev formula denoted by IB), the linear (VL),
nonlinear (VNL), and derivative (VD) time-odd terms in the isoscalar-vector channel.
The calculation is performed using the cranking RMF framework with the PC-F1 interaction,
and the cranking frequency is $\Omega=0.001$~MeV.}
\begin{center}
\begin{tabular}{cccc}
\hline\hline
${\cal I}^{\rm IB}_{x} $  & ${\cal I}^{\rm VL}_{x} $  &  ${\cal I}^{\rm VNL}_{x} $  &  ${\cal I}^{\rm VD}_{x} $  \\
\hline
   55.53   & 18.99  & -0.31   & 1.44   \\
\hline\hline
\end{tabular}
\end{center}
\end{table}

To verify our assumption that the time-odd part of the residual interaction can be approximated
by the linear vector  term, we have analyzed the contributions of the different
time-odd terms to the moments of inertia by performing a self-consistent
cranking calculation (see Refs. \cite{VALR.05,Zhao.10} and references cited therein).
In the cranking framework there are two types of moments of inertia: the kinematic (or static) moment
of inertia $\mathcal{J}^{(1)}$, and the dynamic moment of inertia $\mathcal{J}^{(2)}$. They are defined
as follows
\begin{equation}
\mathcal{J}^{(1)}(\Omega) = \frac{J}{\Omega}, \quad\quad\quad
\mathcal{J}^{(2)}(\Omega) = \frac{dJ}{d\Omega} .
\end{equation}
In a self-consistent calculation with very small vlaues of the rotational frequency, ${\cal J}^{(2)}(\Omega)$ is
identical to the Thouless-Valatin moment of inertia, the linear response to the external Coriolis field.
At the band-head in even-even nuclei the two quantities ${\cal J}^{(1)}$ and ${\cal J}^{(2)}$ coincide
and we use in the figures the character ${\cal I}$ for this quantity.
Calculations that neglect the time-odd fields and take into account only the Coriolis operator
$\Omega \hat{J}_x$ in the Dirac equation, underestimate the empirical moments of inertia
by roughly $30$\%~\cite{KR.89}.
As an illustrative example, in Fig.~\ref{fig:Sm154-cranking} we plot the dynamic moment of inertia for the
ground state band in $^{154}$Sm. By including only the linear time-odd term (VL) in isoscalar-vector channel
the moment of inertia is enhanced by $34$\%, while the remaining two contributions: the non-linear term (VNL)
and the derivative term (VD) yield less than $3$\%. The results are summarized in Tab.~\ref{Tab:Sm154-craning},
where we list the contributions of the linear vector, nonlinear vector and vector derivative terms to the moment
of inertia. Thus in the remaining calculations the model includes only the linear time-odd term.

\begin{figure}[htb!]
\centering
\includegraphics[scale=0.4]{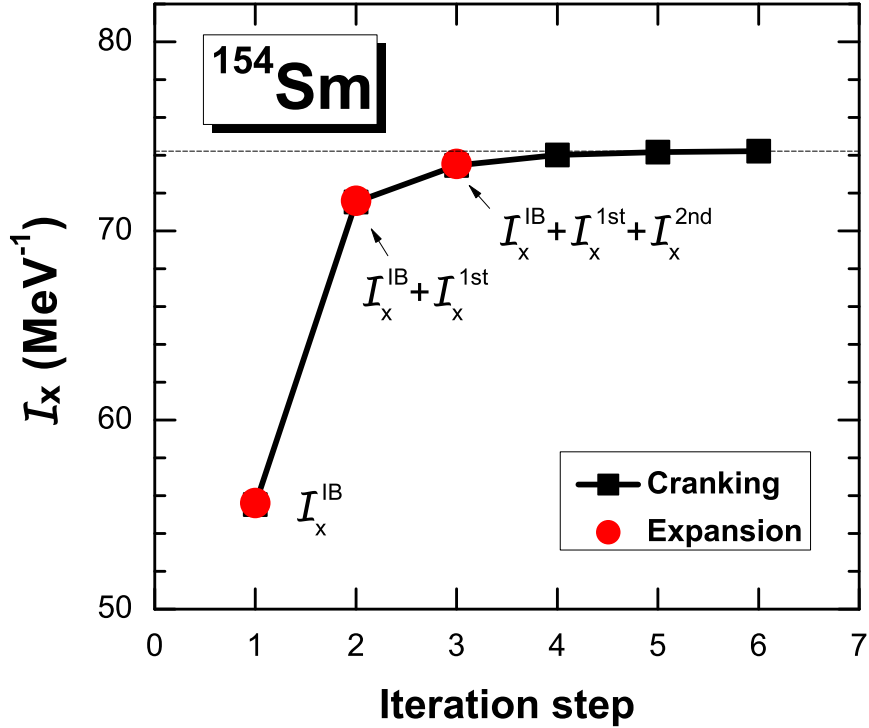}
\caption{\label{fig:Sm154-LRE}(Color online)
The moment of inertia ${\cal I}_x$ of $^{154}$Sm computed at the zeroth-, first-, and second-order
in the expansion Eq.~(\ref{eq:M-expansion}) based on the PC-F1 density functional, and compared
with the values obtained at the corresponding iteration steps in the cranking RMF based on the
same density functional.}
\end{figure}

To estimate the convergence of the expansion formula (\ref{eq:M-expansion}),
we have used it to calculate the moment of inertia ${\cal I}_x$ for the $^{154}$Sm isotope, in
comparison with the exact Thouless-Valatin moment of inertia computed with the cranking code.
In Fig.~\ref{fig:Sm154-LRE} the latter is compared with the zeroth-, first-, and second-order
in the expansion Eq.~(\ref{eq:M-expansion}). Moreover, we also display
the cranking results for each iteration step in the self-consistent cranking calculation,
starting from the stationary solution without the cranking term (cf. Appendix).
As one expects, the moment of inertia obtained after the first iteration
is equal to the Inglis-Belyaev moment of inertia. The next two iterations are compared to the first
and second order in the expansion formula (\ref{eq:M-expansion}). We note that the
values obtained after the second and third iteration steps are in complete agreement with the
first- and the second-order corrections, respectively. In the Appendix we demonstrate that these
values have to be identical, thus the results displayed in Fig.~\ref{fig:Sm154-LRE} provide
a crucial test for the numerical implementation of the expansion Eq.~(\ref{eq:M-expansion}).
Further iteration steps contribute to the value of the moment of inertia by less than 1\%, that is,
the convergence is quite rapid. We also emphasize that it is necessary to include the contributions
from the negative-energy single-nucleon Dirac states in the calculation of the matrix $\mathcal{M}$.
Omitting the negative energy states leads to a significant overestimation of the second-order correction
to the moment of inertia.

\begin{figure}[htb!]
\centering
\includegraphics[scale=0.35]{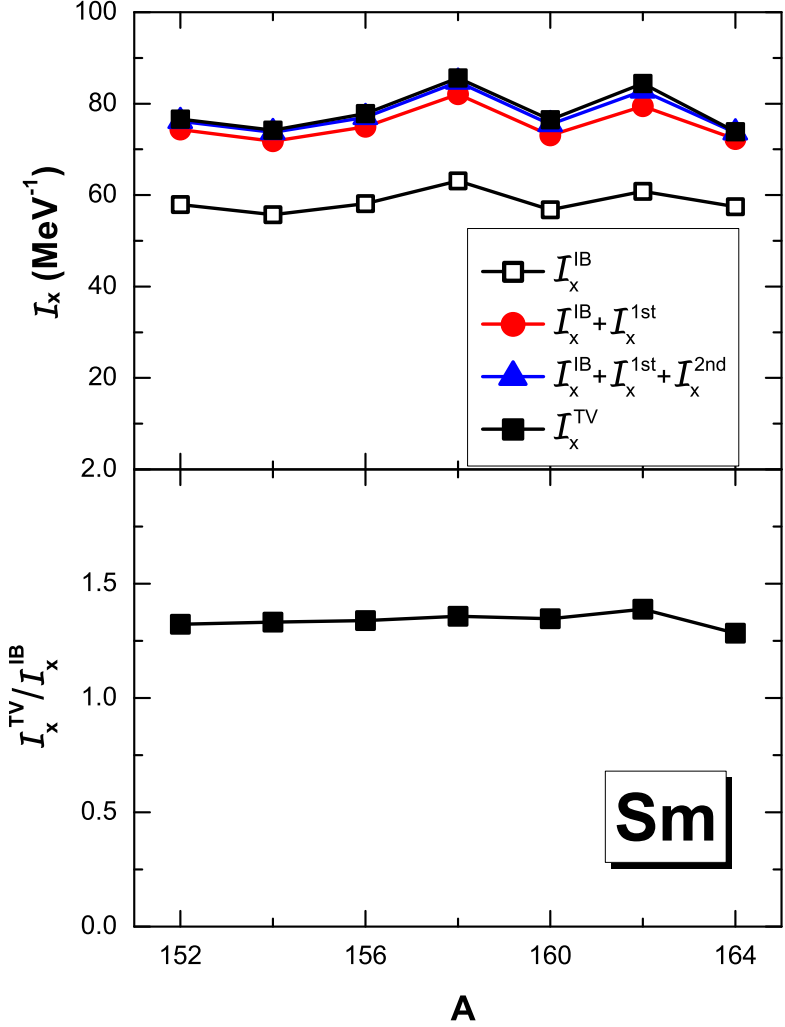}
\caption{\label{fig:Sm-conv}(Color online)
The moments of inertia ${\cal I}_x$ of $^{152-164}$Sm.
The values computed at the zeroth-, first-, and second-order
in the expansion Eq.~(\ref{eq:M-expansion}) are compared
with the exact Thouless-Valatin moment of inertia calculated using the RMF
cranking model (upper panel). For the same nuclei the
ratio of the Thouless-Valatin and Inglis-Belyaev moments of inertia (lower panel).}
\end{figure}

Finally, in Fig.~\ref{fig:Sm-conv} we display the moments of inertia ${\cal I}_x$ for the sequence of
even-even isotopes $^{152-164}$Sm.
The values computed at the zeroth-, first-, and second-order
in the expansion Eq.~(\ref{eq:M-expansion}) are compared
with the exact Thouless-Valatin moment of inertia calculated using the RMF cranking model.
Through the whole isotopic chain
the expansion method, truncated to second order, yields values very close to the exact
Thouless-Valatin moments of inertia, with the relative deviation $\sim 1$\%.
We note that the enhancement of the moment of inertia in comparison to the Inglis-Belyaev
value ranges of $1.28 \sim 1.39$.

\section{\label{sec-summary} Summary and outlook}
Starting from the adiabatic time-dependent Hartree-Fock theory, we have introduced an
efficient approximate method to calculate the Thouless-Valatin moments of inertia for the nuclear system.
The method is based on the fact that the expansion of the inertial parameters
converges rapidly because the matrix elements of the time-odd components of the residual
interaction are usually small in comparison to the $ph$-energies. This approximation is
computationally much less demanding than the full ATDHF calculation,
yet it retains high accuracy of the order of $1$\%. The accuracy of this method has
been verified by comparing the results to the exact Thouless-Valatin rotational moments of inertia
calculated within the cranking model.

One might, of course, encounter problems in regions of level crossings, where the $ph$ energies
are no longer necessarily small compared to the matrix elements of the residual interaction
$\mathcal{V}$. In that case the matrix $\mathcal{M}^{-1}$ in Eq. (\ref{eq:E18})
has to be decomposed in a different way as, for instance, by shifting the diagonal elements of
$\mathcal{V}$ to $\mathcal{M}^{-1}_0$, or by adding and subtracting complex diagonal elements.

The present study has been limited to the rotational moments of inertia. In future investigations
we plan also the calculations of vibrational masses. In this case the momentum operator $P$ is not known a priori. Several
ways have been proposed in the literature~\cite{YLQ.99,DS.81,BSD.11}
to attack this problem. We hope to solve this problem by a similar expansion as in Eq. (\ref{eq:M-expansion}).
Of course this method can also be used with different density functionals by simply replacing the time-odd residual interaction.
Pairing correlations can be included by expanding the inverse of the QRPA matrix instead of the RPA matrix.
Work in this direction is already in progress.

\begin{acknowledgements}
We thank J. Dobaczewski, H. Z. Liang, and P. W. Zhao for very helpful discussions.
This work was supported in part by the Major State 973 Program 2007CB815000,
the NSFC under Grant Nos. 10975008, 10947013, 11105110, and 11105111,
the Southwest University Initial Research
Foundation Grant to Doctor (Nos. SWU110039, SWU109011), the Fundamental
Research Funds for the Central Universities (XDJK2010B007 and XDJK2011B002),
the MZOS - project 1191005-1010, and the DFG cluster of excellence \textquotedblleft Origin and
Structure of the Universe\textquotedblright\ (www.universe-cluster.de).
The work of J.M., T.N., and D.V. was supported in part by
the Chinese-Croatian project "Nuclear structure and astrophysical applications".
T. N. and Z. P. Li acknowledge support by the Croatian National Foundation for
Science.
\end{acknowledgements}
\section*{\label{App}Appendix: Iterative solution of the cranking equation}
In this appendix it is demonstrated that the moment of inertia calculated at each step of the iterative
solution of the cranking equation, coincides with the corresponding order of the expansion
introduced in Sec.~\ref{sec-theoretical}. We assume that the cranking frequency in the equation of motion
\begin{equation}
\left[  h(\rho)-\Omega j_x,\rho\right]  = 0 \;,
\end{equation}
is an infinitesimal quantity, that is, second and higher order terms in $\Omega$ can be safely neglected.
As the initial point we choose the self-consistent solution $\rho_0$ for frequency $\Omega=0$.
The corresponding equation of motion reads:
\begin{equation}
[h_0,\rho_0]=0.
\end{equation}
In the first step of the iteration we diagonalize the operator $h_0-\Omega j_x$, and compute the
density $\rho_0 + \delta \rho_0$ determined by the following relation:
\begin{equation}
\left[  h_{0},\delta\rho_{0}\right]  =\Omega\left[j_x,\rho_{0}\right].
\label{E34}
\end{equation}
In the basis which diagonalizes $h_0$, the only non-vanishing matrix elements of
$\delta \rho_0$ are $ph$ and $hp$. Using the definition of the matrix $\mathcal{M}_0$
Eq.~(\ref{eq:M_0}), we obtain
 \begin{equation}
 \left( \begin{array}{c} \delta \rho_0 \\ \delta \rho^*_0 \end{array} \right)
  = \Omega \mathcal{M}_0 \left( \begin{array}{c}  j_x \\  j^*_x \end{array} \right).
 \end{equation}
 In the following the shorthand notation is used:
 \begin{equation}
 \delta \tilde{\rho}_0 \equiv  \left( \begin{array}{c} \delta \rho_0 \\ \delta \rho^*_0 \end{array} \right) \quad \textnormal{and}
 \quad
 \tilde{j}_x \equiv \left( \begin{array}{c}  j_x \\  j^*_x \end{array} \right)\;,
 \end{equation}
 that is, $\delta \tilde{\rho}_0 = \Omega \mathcal{M}_0 \tilde{j}_x$.
After the first iteration we obtain the Inglis-Belyaev moment of inertia:
\begin{equation}
I_0 =\frac{1}{\Omega} \textnormal{Tr}(j_x\delta\rho_0) = \frac{1}{\Omega} \left( \begin{array}{cc} j_x^* & j_x \end{array}\right)
 \left( \begin{array}{c} \delta \rho_0 \\ \delta \rho^*_0 \end{array} \right)
 =\tilde{j}_x^\dagger \mathcal{M}_0 \tilde{j}_x \equiv I^{\textnormal{IB}}.
\end{equation}
In the second iteration we diagonalize the operator:
\begin{equation}
h_1 - \Omega j_x = h(\rho_0 + \delta \rho_0 ) - \Omega j_x = h_0 + \mathcal{V}\delta\rho_0
-\Omega{j}_x  = h_0 - \Omega\left({j}_x - \mathcal{V}\mathcal{M}_0\tilde{j}_x\right),
\end{equation}
where $\mathcal{V}\mathcal{M}_0\tilde{j}_x$ denotes the matrix
\begin{equation}
\left(\mathcal{V}\mathcal{M}_0\tilde{j}_x\right)_{ab}=
\sum_{php'h'} \mathcal{V}_{ahbp}\mathcal{M}_{0php'h'}(j_x)_{p'h'}+\mathcal{V}_{apbh}\mathcal{M}_{0hph'p'}{(j_x)}^*_{p'h'} \;.
\label{E39}
\end{equation}
The density $\rho_0 + \delta\rho_1$ is the solution of the equation of motion
\begin{equation}
 [h_0 - \Omega\left(j_x - \mathcal{V}\mathcal{M}_0\tilde{j}_x\right), \rho_0+\delta \rho_1] =0 \;.
\end{equation}
Again, $\delta\rho_1$ has only $ph$-matrix elements and, therefore, we need only these elements
of the matrix (\ref{E39}) and find:
\begin{equation}
\delta \tilde{\rho}_1 = \Omega \mathcal{M}_0 \left( \openone - \mathcal{V}\mathcal{M}_0\right)
\tilde{j}_x.
\end{equation}
The moment of inertia obtained in the second iteration coincides with that defined by Eq. (\ref{E23})
\begin{equation}
I_2 = \frac{1}{\Omega}\textnormal{Tr}(j_x \delta \rho_1) =
\left( \begin{array}{cc} j_x^* & j_x \end{array}\right)
 \left( \begin{array}{c} \delta \rho_1 \\ \delta \rho^*_1 \end{array} \right)
 =\tilde{j}_x^\dagger\mathcal{M}_0 \left( \openone - \mathcal{V}\mathcal{M}_0\right)
\tilde{j}_x \;.
\end{equation}
Obviously this can be continued, and finally we obtain the expansion for the full moment of inertia
\begin{equation}
I = \tilde{j}^\dag_x\left(\mathcal{M}_{0}-\mathcal{M}_{0}\mathcal{V M}_{0}+
\mathcal{M}_{0}\mathcal{VM}_{0}\mathcal{VM}_{0}-\dots\right)\tilde{j}_x \;,
\end{equation}
which is equivalent to the expansion of the matrix $\mathcal{M}$ in Eq. (\ref{eq:M-expansion}).
%


\end{document}